\documentclass[12pt]{article}
\usepackage{epsfig}
\usepackage{a4,isolatin1}
\usepackage{amsmath,amsfonts,latexsym, amssymb}

\newtheorem{satz}{Theorem}[section]
\newtheorem{defi}[satz]{Definition}

\newtheorem{bem}[satz]{Remark}
\newtheorem{lemma}[satz]{Lemma}
\newtheorem{koro}[satz]{Corollary}
\newtheorem{bsp}[satz]{Example}

\newtheorem{conclusion}[satz]{Conclusion}
\newtheorem{ob}[satz]{Observation}

\newtheorem{problem}[satz]{Problem}

\newcommand{\mcal}{\mathcal}

\newcommand{\tit}{\textit}

\newcommand{\C}{\mathbb{C}}
\newcommand{\N}{\mathbb{N}}

\newcommand{\Z}{\mathbb{Z}}

\newcommand{\bewende}{$ \hfill \Box $}

\newcommand{\Hia}{{\it H_0}}

\newcommand{\Fc}{\cal F}

\newcommand{\dast}{d^{\ast}}

\begin{document}
\thispagestyle{empty}
\begin{center}
\vspace*{1.0cm}

{\LARGE{\bf Graph-Laplacians and Dirac Operators  on (Infinite) Graphs
\\and the Calculation of the Connes-Distance-Functional}} 

\vskip 1.5cm

{\large {\bf Manfred Requardt }} 

\vskip 0.5 cm 

Institut f\"ur Theoretische Physik \\ 
Universit\"at G\"ottingen \\ 
Bunsenstrasse 9 \\ 
37073 G\"ottingen \quad Germany\\
(E-mail: requardt@theorie.physik.uni-goettingen.de)

\end{center}

\vspace{1 cm}

\begin{abstract}
  We develop a graph-Hilbert-space framework, inspired by
  non-commutative geometry, on (infinite) graphs and use it to study
  spectral properies of \tit{graph-Laplacians} and so-called
  \tit{graph-Dirac-operators}. Putting the various pieces together we
  define a {\it spectral triplet} sharing most (if not all, depending
  on the particular graph model) of the properties of what Connes
  calls a \tit{spectral triple}. With the help of this scheme we
  derive an explicit expression for the {\it Connes-distance function}
  on general graphs and prove both a variety of apriori estimates for
  it and calculate it for certain examples of graphs. As a possibly
  interesting aside, we show that the natural setting of approaching
  such problems may be the framework of \tit{(non-)linear programming}
  or \tit{optimization}. We compare our results (arrived at within our
  particular framework) with the results of other authors and show
  that the seeming differences depend on the use of different
  graph-geometries and/or Dirac operators.
\end{abstract} \newpage
\setcounter{page}{1}

\section{Introduction}
We recently embarked on a programme to reconstruct continuous physics
and/or mathematics from an underlying more primordial and basically
discrete theory living on the Planck-scale
(cf. \cite{1},\cite{2},\cite{quantum}). As sort of a ``spin-off''
various problems of a more mathematical and technical flavor emerged
which have an interest of their own. \tit{Discrete differential
  geometric} concepts were dealt with in \cite{1}, the theory of
\tit{random graphs} was a central theme of \cite{2}, topics of
\tit{dimension theory} and \tit{fractal geometry} were addressed in
\cite{3}.

If one wants to recover the usual \tit{(differential) operators} of
continuum physics and mathematics by some sort of limiting process
from their discrete protoforms, living on a relatively disordered
discrete background like, say, a network, one has, in a first step, to
study their discrete counterparts. This will be our main theme in the
following with particular emphasis on discrete \tit{Laplacians} and
\tit{Dirac-operators} on general \tit{graphs}.

We note in passing that \tit{functional analysis on graphs} is both of
interest in pure and applied mathematics and also in various fields of
(mathematical) physics. For one, discrete systems have an increasing
interest of their own or serve as easier to analyse prototypes of
their continuum counterparts. To mention a few fields of applications:
\tit{graph theory in general, analysis on (discrete) manifolds,
  lattice or discretized versions of physical models in statistical
  mechanics and quantum field theory, non-commutative geometry,
  networks, fractal geometry} etc. From the vast and widely scattered
literature we mention (possibly) very few sources which were of
relevance for our own motivation or we came across recently (after
finishing a first draft): \cite{Harary} to \cite{Rieffel1},\cite{4}
and \cite{Landi}, paper \cite{Iochum} appeared after finishing our
first draft, some more literature like e.g. \cite{Davies} was pointed
out to us by Mueller-Hoissen; the possible relevance of references
\cite{Colin} to \cite{Schrader} were brought to our notice by some
unknown referee, \cite{Novikov} was also discovered by us only
recently. Last, but not least, there is the vast field of
\tit{discretized quantum gravitiy} (see e.g. \cite{Loll} or
\cite{Ambjorn}).  All this shows that the sort of discrete functional
analysis we are dealing with in the following, is presently a very active field
with a lot of different applications.

We use the graph-Hilbert-space machinery, developed in the first part
of our paper to investigate the spectral properties of \tit{graph
  Laplacians} and \tit{Dirac operators}. In a next step we study and
test concepts and ideas, which arose in the framework of
\tit{non-commutative geometry}. As we (and others) showed in preceding
papers, networks and graphs may (or even should) be understood as
examples of \tit{non-commutative} spaces. A currently interesting
topic in this field is the investigation of certain \tit{distance
  functionals} on ``nasty'' or \tit{non-standard} spaces and their
mathematical or physical ``naturalness''.  Graphs carry, on the one
hand, a natural {\it metric structure} given by a {\it distance
  function} $d(x,y)$, with $x,y$ two {\it nodes} of the graph (see the
following sections). This fact was already employed by us in e.g.
\cite{3} to develop dimensional concepts on graphs. Having Connes'
concept of distance in noncommutative geometry in mind (cf. chapt. VI
of \cite{4}), it is a natural question to try to compute it in model
systems, which means in our context: arbitrary graphs, and compare it
with the already existing notion of graph distance mentioned above.
(We note in passing that the calculation of the Connes distance for
general graphs turns out to be surprisingly complex and leads to
perhaps unexpected connections to fields of mathematics like e.g.
\tit{(non-)linear programming} or \tit{optimization}; see the last section).

Therefore, as one of many possible applications we construct a
protoform of what Connes calls a {\it spectral triple}, that is, a
Hilbert space structure , a corresponding representation of a certain
(function) algebra and a (in our framework) natural candidate for a
so-called {\it Dirac operator} (not to be confused with the ordinary
Dirac operator of the Dirac equation), which encodes certain
properties of the {\it graph geometry}. This will be done in section
4.

In the last section, which deals with the distance concept deriving
from this spectral triplet (as we like to call it), we will give this
notion a closer inspection as far as graphs and similar spaces are
concerned. In this connection some recent work should be mentioned, in
which Connes' distance function was analyzed in certain simple models
like e.g. one-dimensional lattices (\cite{5}-\cite{7}). These papers
already show that it is a touchy business to isolate ``the''
appropriate Dirac operator (after all, different Dirac operators are
expected to lead to different geometries!)  and that it is perhaps
worthwhile to scrutinize the whole topic in a more systematic way. We
show in particular that one may choose different Dirac-operators on
graphs (or rather, different types of graphs over the same node set) which may lead to different results for e.g. the corresponding
Connes-distance. 

The problem of finding suitable metrics on ``non-standard'' spaces is
a particularly interesting research topic of its own, presently
pursued by quite a few people (see the beautiful paper by Rieffel,
\cite{Rieffel1} and the references mentioned therein). Another earlier
source is e.g. \cite{Davies}. As to this latter paper we would like to
remark that, while much of the framework is different from ours, there
is, on the other side, a small overlap as far as some technical
notions and results are concerned (after an appropriate translation of
the respective technical notions and definitions). To give an example:
While the definition of Dirac operators is different, some of the
operator norms and metrics, being calculated, turn out to be
identical to ours. This suggests a more careful comparison of the underlying
conceptual ideas which we plan to give elsewhere. We presently extend
this investigation of metric structures to \tit{lump spaces} and
\tit{probabilistic metric spaces} (see \cite{requroy} and
\cite{schweizer}) in the general context of quantum gravity (cf. also \cite{2}).

Our own approach provides a \tit{systematic} recipe to calculate the
{\it Connes distance} in the most general cases of graphs and exhibits
its role as a non-trivial constraint on certain function classes on
graphs. We prove various rigorous a priori estimates and show how the
constraints have to be dealt with in
several examples.\\[0.3cm]
Remark: (For possible reasons of priority) we would like to mention
that many of our results can already be found in a (preliminary form) in
an earlier draft version (\cite{Graphs}).

\section{A brief Survey of Differential Calculus on\\ Graphs}

The following is a brief survey of certain concepts and tools needed
in the further analysis. While our framework may deviate at various
places from the ordinary one, employed in e.g. \tit{algebraic graph
  theory}, this is mainly done for reasons of greater mathematical
flexibility and generality and, on the other side, possible physical
applications (a case in point being the analysis of \tit{non-commutative
  spaces}). Some more motivations are provided in \cite{1} and
\cite{2}). We begin with the introduction of some graph theoretical
concepts. We would however like to mention, that it is not our
intention to cover any appreciable amount of the close
interrelationship between graph spectra and graph characteristics (as
has e.g. been done in \cite{Colin}; our main emphasis lies on
providing various Hilbert-space-techniques).
\begin{defi}[Simple Locally Finite (Un)directed Graph]\hfill
\begin{enumerate}
\item We write the {\em simple} graph as $G:=(\mcal{V},E)$ where $\mcal{V}$ is the
  countable set of nodes $\{n_i\}$ (or vertices) and $E$ the set of
  bonds (edges). The graph is called simple if there do not exist
  elementary {\em loops} and {\em multiple edges}, in other words: each
  existing bond connects two different nodes and there exists at most
  one bond between two nodes. (We could of course also discuss more
  general graphs). Furthermore, for simplicity, we assume the graph to
  be connected, i.e. two arbitrary nodes can be connected by a
  sequence of consecutive bonds called an {\em edge sequence} or {\em walk}. A
  minimal edge sequence, that is one with each intermediate node
  occurring only once, is called a {\em path} (note that these definitions
  may change from author to author).
\item We assume the graph to be {\em locally finite}, that is, each
  node is incident with only a finite number of bonds. Sometimes it is
  useful to make the stronger assumption that this {\em vertex
  degree}, $v_i$, (number of bonds being incident with $n_i$), is
  globally bounded away from $\infty$.
\item One can give the edges both an {\em orientation} and a {\em
    direction} (these two, in our view, slightly different geometric concepts are
    frequently intermixed in the literature). In our context we adopt
    the following convention: If two nodes $n_i,n_k$ are connected by
    a bond, we interpret this as follows: There exists a {\em
    directed bond}, $d_{ik}$, pointing from $n_i$ to $n_k$ and a {\em
    directed bond}, $d_{ki}$, pointing in the opposite direction. In
    an algebraic sense, which will become clear below (for more
    details see also \cite{1}), we call
    their {\em superposition}
\begin{equation} b_{ik}:= d_{ik}-d_{ki}= -b_{ki}\end{equation}
the corresponding {\em oriented bond} (for obvious reasons; the
    directions are fixed while the orientation can change its sign). In
    a sense the above reflects the equivalence of an {\it undirected
    graph} with a {\em directed multi-graph} having two directed
    bonds pointing in opposite directions for each undirected bond.
\end{enumerate}
\end{defi}
This way of algebraic implementation of geometric structures allows us
to treat in principle all kinds of graphs on essentially the same
footing. That is, it also applies to, say, graphs with only one
directed edge being existent between two nodes. On the other side, an
orientation should exist also for undirected graphs. As an aside, we
remark that, on the one side, our generators, $b_{ik}$, correspond to
the oriented pairs of nodes, $(i,k)$ in e.g. \cite{Colin}, on the
other hand, our $d_{ik}$ correspond to the oriented pairs, $(i,k)$, in
\cite{Davies}. One sees from this that the conventions are far from
being unique and that a certain unification may be desirable.

We now take the elementary building blocks $\{n_i\}$ and $\{d_{ik}\}$
as basis elements  of a certain hierarchy of vector spaces over, say,
$\C$ with scalar product
\begin{equation}
(n_i|n_k)=\delta_{ik}\quad(d_{ik}|d_{lm})=\delta_{il}\cdot\delta_{km}\end{equation}
\begin{defi}[Vertex-, Edge-Space]
  The vector spaces (or modules) $C_0$, $C^a_1$ ($a$ for
  antisymmetric) and $C_1$ consist of the finite sums
\begin{equation} f:=\sum f_in_i\quad g:=\sum
  g_{ik}d_{ik}\quad\mbox{with}\quad g_{ik}=-g_{ki}\quad\mbox{and}\quad
 g':=\sum g_{ik}d_{ik} \end{equation}
$f_i,g_{ik}$ ranging over a certain given field like e.g. $\C$
or ring like e.g. $\Z$ in case of a module. Evidently we have
$C^a_1\subset C_1$.
\end{defi}
 These spaces can be easily completed to {\it Hilbert spaces} by assuming
\begin{equation} \sum |f_i|^2<\infty\quad \sum |g_{ik}|^2<\infty\end{equation}
if one chooses e.g. the field $\C$ (see the next section).
Furthermore, one can continue this row of vector spaces in  ways which
are common practice in, say, {\em algebraic topology} ( see \cite{1}
sections 3.1 and 3.2).
In this context they are frequently called {\em chain complexes} (see
also \cite{Novikov}).
Evidently the above vector spaces could as well be viewed as {\em
  discrete function spaces} over the {\em node-, bond set} with
$n_i,d_{ik}$ now representing the elementary {\em indicator
  functions}.

In the same spirit we can now introduce two linear maps between
$C_0,C_1$ called for obvious reasons \tit{boundary-} and
\tit{coboundary map}. On the basis elements they act as follows:
\begin{defi}[(Co)boundary Operator]
\begin{equation} \delta:\; d_{ik}\to n_k\quad\text{hence}\quad b_{ik}\to
n_k-n_i\end{equation}
\begin{equation} d:\; n_i\to \sum_k(d_{ki}-d_{ik})=\sum_k b_{ki}\end{equation}
and linearly extended. That is, $\delta$ maps the directed bonds
$d_{ik}$ onto the terminal node  and $b_{ik}$ onto its (oriented) {\em boundary},
while $d$ maps the node $n_i$ onto the sum of the {\em ingoing} directed
bonds minus the sum of the {\em outgoing} directed bonds or on the sum of
{\em oriented} {\em ingoing} bonds $b_{ki}$.
\end{defi}

The following results show, that these definitions lead in fact to a
kind of \tit{discrete differential calculus} on $C_0,C_1$.
\begin{ob}[Discrete Differential Forms]
From the above it follows that
\begin{equation} df=d(\sum f_in_i)=\sum_{k,i}(f_k-f_i)d_{ik}\end{equation}
\end{ob}
Combining now the operators $\delta$ and $d$, we can construct, what
is called the \tit{canonical graph Laplacian}. On the vertex space it reads:
\begin{ob}[Graph Laplacian]
\begin{equation} \delta df=-\sum_i(\sum_k f_k-v_i\cdot
  f_i)n_i=-\sum_i(\sum_k(f_k-f_i))n_i=:-\Delta f \end{equation} where
$v_i$ denotes the {\em node degree} or {\em valency} defined above and
the $k$-sum extends over the nodes adjacent to $n_i$.
\end{ob}
Note that there exist several variants in the literature (see e.g.
\cite{Colin} or \cite{9}). Furthermore, many mathematicians
employ a different sign-convention. We stick in the following to the
convention being in use in the mathematical-physcis literature
where $-\Delta$ is the positive(!) operator.

This \tit{graph Laplacian} is intimately connected with yet another
important object, employed by graph-theorists, i.e. the {\it adjacency matrix} of a graph.
\begin{defi}[Adjacency Matrix] The entries $a_{ik}$ of the
{\em adjacency matrix} $A$ have the value one if the nodes $n_i,n_k$
are connected by a bond and are zero elsewhere. If the graph is {\em
  undirected} (but orientable; the case we mainly discuss), the
relation between $n_i,n_k$ is {\em symmetric}, i.e.
\begin{equation} a_{ik}=1\quad\Rightarrow\quad a_{ki}=1\quad\text{etc.}\end{equation}
This has the obvious consequence that in case the graph is {\em
  simple} and {\em undirected}, $A$ is a
symmetric matrix with zero diagonal elements.
\end{defi}
Remark: More general $A$'s occur if more general graphs are
admitted (e.g. general multigraphs).
\begin{ob}: With our definition of $\Delta$ it holds:
\begin{equation} \Delta=A-V\end{equation}
where $V$ is the diagonal {\em degree matrix}, having $v_i$ as
diagonal entries.
\end{ob}
(Note that the other sign-convention would lead to $\Delta=V-A$).\\[0.3cm]
Proof: As we have not yet introduced the full Hilbert space machinery (which
we will introduce in the next section), the proof has to be understood, for the time being,
in an algebraic way. We then have:
\begin{equation} Af=A(\sum
f_in_i)=\sum_if_i(\sum_{k-i}n_k)=\sum_i(\sum_{k-i}f_k)n_i\end{equation}
\begin{equation} Vf=\sum_i(v_if_i)n_i\end{equation}
hence the result.\bewende \\[0.3cm]
(Here and in the following we use the abbreviation $k-i$ if
the nodes $n_k,n_i$ are connected by a bond, the summation always
extending over the first variable). 

As we already remarked above, our approach to functional analysis on
graphs is perhaps a little bit different compared with the usual one.
It seems therefore to be appropriate to exhibit some of the conceptual
differences as compared to the more traditional framework, as e.g.
developed in the beautiful monographs \cite{8} or \cite{9}, by briefly
discussing the following (however only minor) point. In general, the graphs
under discussion may be directed or undirected. In the traditional
approach the edges are typically independently labelled of the nodes
and the corresponding edge space, denoted in this case for the time
being by $\hat{C}_1$, is built over this edge set. In contrast to that
habit we found it useful to label the occurring edges as
$d_{ik},b_{ik}$ with $b_{ik}=-b_{ki}$ which leads in our view to a
more flexible discrete calculus and, among other things, to a natural
\tit{Dirac operator} on graphs (see below).

More or less related to our operators $d,d^*$ ($d^*$ the adjoint of
$d$; see the next section) are now the so-called
\tit{incidence matrix}, $B$, and its adjoint in the traditional
approach which relate the edges with the nodes. To do this, the edges
are given an adhoc orientation, denoting one vertex arbitrarily as
\tit{initial point}, the other as \tit{end point}. With the $n$
labelled vertices, $n_i$ and $m$ labelled edges, $e_j$ $(B_{ij})$ has
the entries
\begin{equation}
\begin{cases}+1 & \text{if $n_i$ is the endpoint of $e_j$}\\
-1 & \text{if $n_i$ is the initial point of $e_j$}\\
0 & \text{otherwise}
\end{cases}
\end{equation}
Evidently $B$ is a mapping from $\hat{C}_1$ to $C_0$ and maps an edge
to the respective difference of end vertex and initial vertex. By the
same token, the transpose, $B^t$ is defined as a map from $C_0$ to
$\hat{C}_1$ and one gets:
\begin{equation}B\cdot B^t=V-A\end{equation}
Note that the above introduced adhoc orientation does not enter in any
end result; on the other hand, our approach is not based on such a
contingent structure.

\section{Some Spectral Analysis and Operator Theory on (Infinite) Graphs}
After these preliminary remarks we now enter the heart of the matter.
Our first task consists of endowing a general graph with a natural
Hilbert space structure on which the various operators constructed in
the following can operate.
(The following analysis is done on {\em
    undirected} graphs, but could be extended to more general but
  less symmetric situations).
\begin{defi}[Hilbert Space] As indicated in the previous section, we extend
  $C_0,C_1^a,C_1$ to the respective Hilbert spaces $H_0,H_1^a\subset H_1$ of
  sequences over the respective ON-bases $\{n_i\},\,\{d_{ik}\}$, that
  is $(n_i|n_k)=\delta_{i,k}$,
  $(d_{ik}|d_{i'k'})=\delta_{ii'}\delta_{kk'}$.

 As $H^a,H$ we take the direct sums:
\begin{equation}  H^a:= H_0\oplus H_1^a\subset H:= H_0\oplus
  H_1\end{equation}
\end{defi}
Note that members of $ H_1^a$ can  be written
\begin{equation} \sum g_{ik}d_{ik}=1/2\sum g_{ik}d_{ik}+1/2\sum
g_{ki}d_{ki}=1/2\sum g_{ik}(d_{ik}-d_{ki})=1/2\sum g_{ik}b_{ik}\end{equation}
Obviously $H^a$ is a subspace of $H$ and we
have
\begin{equation} (b_{ik}|b_{ik})=2\end{equation}
i.e. the $b_{ik}$ are not(!) normalized if the $d_{ik}$ are. We could
of course enforce this but then a factor two would enter
elsewhere.

With these definitions it is now possible to define the maps
$d,\,\delta$ as true operators between these Hilbert (sub)spaces. To
avoid domain problems we assume from now on that the {\it node degree}
$v(n_i)$ is {\it uniformly bounded} on the graph $G$, i.e.
\begin{equation} v_i\le v_{max}<\infty\end{equation}
\begin{ob}\label{relations} We have the following relations
\begin{equation} d:\;H_0\to H^a_1\subset H_1,\quad\delta:H_1\to H_0\end{equation}
$d_{1,2}$ with
\begin{equation} d_{1,2}:\;n_i\to\sum d_{ki}\; ,\;\sum d_{ik}\end{equation}
respectively and linearly extended, are linear operators from $H_0\to H_1$
and we have
\begin{equation} d=d_1-d_2\end{equation}
Similarly we may define $\delta=:\delta_1$ and $\delta_2$ via:
\begin{equation} \delta_{1,2}:\;d_{ik}\to n_k,\,n_i\end{equation}
\end{ob}

It is remarkable (but actually not surprising) that $v_i\le v_{max}$ implies that all the above
operators are {\it bounded} (in contrast to their continuous counterparts, which are typically unbounded). Taking this for granted at
the moment, there are no domain problems and a straightforward
analysis yields the following relations:
\begin{ob}\hfill \label{newrelations} 
\begin{enumerate}
\item The adjoint $d^{\ast}$ of $d$ with respect to
the spaces $H_0,H^a_1$ is $2\delta$
\item On the other side we have for the natural extension of
  $d,\delta$ to the larger space 
$H_1$ (cf. the definitions in Observation \ref{relations}):
\begin{equation} \delta_1=(d_1)^{\ast}\;,\;\delta_2=(d_2)^{\ast}\end{equation}
hence 
\begin{equation} (\delta_1-\delta_2)=(d_1-d_2)^{\ast}=d^{\ast}\neq 2\delta=2\delta_1\end{equation}
\item Furthermore we have
\begin{equation}d_1^*\cdot d_1=\delta_1\cdot d_1=d_2^*\cdot
  d_2=V:\;n_i\to v_in_i\end{equation}
\begin{equation}d_1^*\cdot d_2=\delta_1\cdot d_2=d_2^*\cdot
  d_1=\delta_2\cdot d_1=A:\; n_i\to\sum_{k-i}n_k\end{equation}
Similar geometric properties of the graph are encoded in the products
coming in reversed order.
\end{enumerate}
\end{ob}

That and how $d,d^*$ encode some more geometric information about the
graph can be seen from the following domain- and range-properties (for
corresponding results in the more traditional approach see also
\cite{8},p.24ff).
\begin{satz}Let the graph be connected and finite, $|\mcal V|=n$, then
\begin{equation}dim(Rg(d^*))=n-1\end{equation}
\begin{equation}dim(Ker(d^*))=\sum_i v_i-(n-1)\end{equation}
With $dim(H_1)=\sum_i v_i$, $dim(H_1^a)=1/2\cdot dim(H_1)$ we have
\begin{equation}codim(Ker(d^*))=dim(Rg(d))=n-1\end{equation}
We see that both $Rg(d^*)$ and $Rg(d)$ have the same dimension $(n-1)$.
\end{satz}
\begin{bem}In case the graph has, say, $c$ components, the above
  results are altered in an obvious way; we have for example
\begin{equation}dim(Rg(d^*))=n-c\end{equation}
\end{bem}
Proof: we first state the general result for bounded operators
\begin{equation}Rg(T^*)=Ker(T)^{\perp}\end{equation}
we then have for $T=d$
\begin{equation}f\in Ker(d)\Rightarrow
  0=d(f)=\sum_{ik}(f_k-f_i)d_{ik}\end{equation}
As the $d_{ik}$ are linearly independent this entails $f_k=f_i$ for
the pairs $(i.k)$ which occur in the sum. Since the graph is connected
we have $f_k=f_i=const$ for all nodes, hence $dim(Ker(d))=1$ and is
spanned by $\sum_i n_i$. this proves the first item.

In a similar way we proceed for $d^*$.
\begin{equation}0=d^*(g)=\sum_i(\sum_k(g_{ki}-g_{ik}))n_i\Rightarrow
  \sum_k(g_{ki}-g_{ik})=0\;\text{for all nodes}\;n_i\end{equation}
In $H_1$ $g_{ik},g_{ki}$ can be independently chosen. We have $n$
linear equations, which are, however, not independent. There is, in
fact, exactly one apriori constraint of the form
\begin{equation}\sum_i(\sum_k(g_{ik}-g_{ki}))=0\end{equation}
Hence, the above yields exactly $n-1$ independent linear equations for
the $\sum v_i$ coefficients. This implies that the subspace, so
defined, has dimension $\sum v_i-(n-1)$. This proves items two and three.
\bewende
\begin{ob}In the literature $Ker(d^*)$ is called (for obvious reasons)
  the {\em cycle subspace} (cf e.g. \cite{8}). On the antisymmetric
  subspace $H_1^a$ we have $d^*=2\delta$ and
  $\delta(b_{ik})=n_k-n_i$. Choosing now a {\em cycle}, given by its
  sequence of consecutive vertices $n_{i_1},\ldots,
  n_{i_k};n_{i_{k+1}}:=n_{i_1}$, we have
\begin{equation}d^*(\sum
  b_{i_li_{l+1}})=2\sum(n_{i_{l+1}}-n_{i_l})=0\end{equation}
that is, vectors of this kind lie in the kernel of $d^*$   
\end{ob}

We will now provide quantitative lower and upper bounds for the respective
norms of the occurring operators. For $d$ we have:
\begin{equation} d:\,\Hia\ni\sum_i
f_in_i\to\sum_if_i(\sum_{k-i}b_{ki})=\sum_{ik}(f_k-f_i)d_{ik}\end{equation}
and it follows for the norm of the rhs:
\begin{multline}
  \|rhs\|^2=\sum_{ik}|(f_k-f_i)|^2=\sum_iv_i\cdot|f_i|^2+\sum_kv_k\cdot|f_k|^2-\sum_{i\neq
    k}(\overline{f_k}f_i+\overline{f_i}f_k)\\
=2\cdot\sum_i
  v_i|f_i|^2-2\cdot\sum_{i\neq k}\overline{f_k}f_i
\end{multline} 
The last expression can hence be written:
\begin{equation} \|df\|^2=2((f|Vf)-(f|Af))=(f|-2\Delta f)\end{equation}
and shows the close relationship of the norm of $d$ with the {\it
  expectation values} of the {\it adjacency} and {\it degree matrix}
respectively the {\it graph Laplacian}. That is, norm estimates for,
say, $d$, derive in a natural manner from the corresponding estimates
for $A$ or $-\Delta$. It follows from the above that we
have:
\begin{ob} 
\begin{equation} \|df\|^2=(f|d^{\ast}df)=(f|-2\Delta f)\end{equation}
i.e.
\begin{equation}
d^{\ast}d\,=\,-2\Delta\quad\text{hence}\quad\|d\|^2=\sup_{\|f\|=1}(f|-2\Delta
f)=\|-2\Delta\|\end{equation}
and
\begin{equation} 0<\sup_{\|f\|=1}(f|-2\Delta f)\leq
2v_{max}+2\sup_{\|f\|=1}|<f|Af>|\end{equation}
i.e.
\begin{equation}\|-\Delta\|\leq v_{max}+\|A\| \end{equation}
\end{ob}
We want to note that we are exclusively using the {\em operator norm}
also for matrices (in contrast to most of the matrix literature),
which is also called the {\it spectral norm}. It is unique in so far
as it coincides with the so-called {\it spectral radius} (cf. e.g.
\cite{11} or \cite{12}), that is
\begin{equation} \|A\|:=\sup\{|\lambda|;\,\lambda\in spectr(A)\}\end{equation}

We now provide upper and lower bounds for the operator norm of the
adjacency matrix, $A$, both in the finite- and infinite-dimensional
case. In \cite{alt} we estimated the upper bound by a method being
different from the following calculation which, being based on
\tit{form-estimates}, is much more direct. The previous proof was
based on the so-called \tit{Gerschgorin-inequality} for finite
matrices and a not entirely straightforward extension to the infinite
dimensional case. We expect this upper bound to be well-known
(cf. e.g. \cite{Colin}, Theorem 2.8 -- \tit{Lemma of Gabber-Galil} --
which is slightly more general). We do not know whether this is also
the case for the lower bound. As such lower bounds are frequently less
straightforward to derive and since, as a byproduct, we develop several
potentially useful techniques, we give our own proof of the lower
bound below. 
\begin{satz}[Norm of $A$] With the adjacency matrix $A$
finite or infinite and a finite $v_{max}$ we have the following
result (a certain fixed labelling of the nodes being assumed):
\begin{equation}\limsup\, n^{-1}\cdot\sum_{i=1}^n v_i\leq \|A\|=\sup\{|\lambda|;\,\lambda\in\;spectr(A)\}
\leq v_{max}\end{equation}
\end{satz}
Proof: In order to prove the upper bound we use a form-estimate
directly for the infinite case. We have
\begin{equation}\label{bound}|(x,Ax)|=|\sum \overline{x_j}a^{ji}x_i|\leq
  \sum_{a^{ji}\neq 0}|x_j|\cdot|x_i|\end{equation}
with $a^{ji}=1$ or $0$. Note that, due to the symmetry of $A$,
each term, $|x_j|\cdot|x_i|$ occurs twice in the above sum. With
\begin{equation}2|x_j|\cdot|x_i|\leq |x_j|^2+|x_i|^2\end{equation}
we have
\begin{equation}\text{rhs of}\;(\ref{bound})\leq
  v_{max}\cdot|x|^2\end{equation}
and hence
\begin{equation}|(x,Ax)|\leq v_{max}(x,x)\end{equation}
Strictly speaking we have the norm-bound up to now only proved for the
above quadratic form. The well-known \tit{Riesz-lemma} associates the
form with a unique bounded operator which is the adjacency matrix we
started with. This proves the first estimate.

To prove the lower bound, we label the nodes or the corresponding
orthonormal basis by $(e_1,e_2,\ldots)$, and introduce the respective
adjacency matrices on the corresponding subspaces, $X_n$, belonging to
the \tit{induced subgraphs}, $G_n$, spanned by $(e_1,\ldots,e_n)$. We
choose a normalized vector, $x_n$, in $X_n$ with all its entries being
$n^{-1/2}$. We then have
\begin{equation}\|A\|=\sup_{\|x\|=1}|(x,Ax)|\geq
  |(x_n,Ax_n)|=|(x_n,A_nx_n)|=n^{-1}\sum_1^n v_i\end{equation}
This proves the theorem.
\begin{lemma}The adjacency matrices, $A_n$, converge strongly to $A$
  and we have in particular $\|A_n\|\nearrow\|A\|$.
\end{lemma}
\begin{bem}To prove strong convergence of operators is of some
  relevance for the limit behavior of {\em spectral properties} of the
  operators $A_n,A$. That is (cf. e.g. \cite{Reed} section VIII.7), we
  have in that case ($A_n,A$ selfadjoint and uniformly bounded)
  $A_n\to A$ in {\em strong resolvent sense}, which implies that the
  spectrum of the limit operator, $A$, cannot suddenly expand, i.e.
\begin{equation}\lambda\in spec(A)\Rightarrow\exists\; \lambda_n\in
  spec(A_n)\; \text{with}\; \lambda_n\to \lambda\end{equation}
and for $a,b\not\in spec_{pp}(A)$
\begin{equation}P_{(a,b)}(A_n)\to
  P_{(a,b)}(A)\;\text{strongly}\end{equation}
\end{bem}
Proof of the Lemma: Strong convergence can be proved as follows.
$A-A_n$ has the matrix representation
\begin{equation} A-A_n=\left( \begin{array}{cc}0 & B_n\\B_n^t & C_n \end{array}
\right) \end{equation}
with
\begin{equation} (A-A_n)x=\left( \begin{array}{c}B_nx'_n \\ 0 \end{array}
\right)+\left( \begin{array}{c}0 \\ B_n^tx_n \end{array}
\right)+\left( \begin{array}{c}0 \\ C_nx'_n \end{array} \right)\end{equation}
where
\begin{equation} x=\left( \begin{array}{c}x_n \\ x'_n \end{array}
\right)\quad\mbox{with}\quad x_n=\sum_1^n
x_ie_i\;,\;x'_n=\sum_{n+1}^{\infty}x_ie_i \end{equation}
(and the $B_n$ not to be confused with the incidence matrices of
section 2)
Multiplying from the left with $x$ we easily establish weak
convergence since
\begin{equation} (x|(A-A_n)x)=(x_n|B_nx'_n)+(x'_n|B_n^tx_n)+(x'_n|C_nx'_n) \end{equation}
and with $n\to\infty$ all the terms on the rhs go to zero, as $\|x'_n\|\to
0$ for $n\to\infty$ since $\|x\|<\infty$ and $B_n,B_n^t$ are again
uniformly bounded. 

To show strong convergence the critical term is $B_n^tx_n$. $B_n^t$
maps the vector $x_n\in X_n$ into $X'_n=X\ominus X_n$, $X_n,X'_n$
living on the node sets $V_n,V-V_n$. As $v_{max}<\infty$ we can find
for each given $n$ a finite, minimal $m_n$ so that all bonds beginning
at nodes of $V_n$ end in $V_{m_n}$, in other words:
\begin{equation}\forall\; n\;\exists\;m_n\ge n\;\text{with}\;B_n^tx_n\in
  X_{m_n}\ominus X_n\end{equation}
or
\begin{equation}B^t_{m_n}x_{m_n}=B^t_{m_n}(x_{m_n}-x_n)\end{equation}
as $B^t_{m_n}x_n=0$ by construction.

The $B^t_n$ are uniformly bounded and $\|x_{m_n}-x_n\|\to 0$ for
$n\to\infty$, hence 
\begin{equation}\|B^t_{m_n}x_{m_n}\|\to
  0\;\text{with}\;n\to\infty\end{equation}
Each $l\in\N$ lies between some $m_n$ and $m_{(n+1)}$ and we have
\begin{equation}B^t_lx_l=B^t_l(x_n+(x_l-x_n))=B^t_l(x_l-x_n)\end{equation}
as $B^t_lx_n=0$ for all $l\ge m_n\ge n$.
\begin{equation}l\to\infty\Rightarrow m_n\to\infty\Rightarrow
  n\to\infty\;\text{hence}\; \|x_l-x_n\|\to 0\end{equation}
which shows that
\begin{equation}s-\lim(B^t_lx_l)=s-\lim(B^t_lx)= 0\end{equation}
\bewende
\begin{bem}A slightly simpler but perhaps less instructive proof can
  be given by exploiting the already established weak convergence
  together with special properties of $A_n,B_n$ etc., yielding
\begin{equation}\lim_n ((A-A_n)x|(A-A_n)x)=-\lim_n (A_nx_n|B_nx'_n)=0\end{equation}
\end{bem}

To prove the monotone convergence of $\|A_n\|$ to $\|A\|$, we proceed
as follows. For the principal minors we have 
\begin{equation} A_n=P_nAP_n\quad\mbox{and}\quad A_n=P_nA_mP_n \end{equation} 
with $P_n$ projecting on the subspace spanned by $e_1,\ldots,e_n$ and
$m\geq n$. Hence
\begin{equation} \|A_n\|\leq\|A\|\quad\mbox{and}\quad \|A_n\|\leq\|A_m\| \end{equation} 
as $ \|P_n\|=1$. From this we see that $\|A_n\|$ is monotonely
increasing with $n\to\infty$ and uniformly bounded by $\|A\|$. In
other words:
\begin{equation} \|A_n\|\to a\leq \|A\|\end{equation}
The equality of $a$ and $\|A\|$ follows then immediately from the
strong convergence of $A_n$ towards $A$. This proves the above lemma.

To test the effectiveness of the upper and lower bounds derived above,
we apply them to a non-trivial model recently discussed in
\cite{Froese}, i.e. the infinite \tit{binary tree} with \tit{root}
$n_0$ where $v_0$ is two and $v_i$ equals three for $i\neq 0$. The
authors show (among other things) that the spectrum consists of the
interval $[-2\sqrt{2},2\sqrt{2}]$, i.e. $\|A\|=2\sqrt{2}$. $v_{max}$
is three, we have to calculate $\limsup 1/n\cdot\sum_1^n v_i$. For
simplicity we choose a subsequence so that $n:=n(N)$ with $N$ denoting
the $N$-th level (consisting of $2^N$ nodes) of the tree starting from
the root $n_0$. Note that in the corresponding \tit{induced subgraph}
$G_N$ the \tit{boundary nodes} sitting in the $N$-th level have only
\tit{node degree} one with respect to $G_N$ but three viewed as nodes
in the full tree.

We then have
\begin{equation}n=\sum_{k=1}^N 2^k\; , \; \sum_{i=0}^{n(N)}
  v_i=2+3\cdot\sum_{k=1}^{N-1}2^k+2^N=3\cdot\sum_{k=0}^N 2^k
-2\cdot2^N - 1\end{equation}
Hence
\begin{equation}\lim_{n(N)} 1/n(N)\sum_{i=0}^{n(N)}=3-2\lim_N
  (\sum_0^N 2^{k-N})^{-1}=2\end{equation}
That is, our genral estimate imply $2\leq\|A\|\leq3$, which is not so bad.
\section{The Spectral Triplet on a general
  (undirected) Graph} Note what we said at the beginning about our
restriction to \tit{undirected} graphs (made, however, only for
convenience!). Furthermore our Dirac operator intertwines node-vectors
and bond-vectors while in other examples it maps node- to
node-functions. Our bond-functions have (in some sense) the character
of \tit{cotangential-vectors}, while in other approaches derivatives
of functions are interpreted as \tit{tangent-vectors}. In our view,
the latter formalism is effective only in certain classes of highly
regular models (like e.g. lattices) where one has kind of global
directions and will become cumbersome for general graphs. We developed
this latter approach a little bit in section 3.3 of \cite{1} and
showed how these
cotangent and tangent vectors can be mapped into each other.\\[0.5cm]
The Hilbert space under discussion in the following is
\begin{equation} H=H_0\oplus H_1\end{equation}
The {\it natural representation} of the function algebra ${\cal F}$
\begin{equation} \{f;f\in {\cal C}_0,\sup_i |f_i|<\infty\}\end{equation}
on $H$ by bounded operators is given by:
\begin{equation} H_0:\;f\cdot f'=\sum f_if'_i\cdot n_i\;\mbox{for}\;f'\in H_0
\end{equation}
\begin{equation} H_1:\;f\cdot\sum g_{ik}d_{ik}:=\sum f_ig_{ik}d_{ik}\end{equation}
From previous work (\cite{1}) we know that ${\cal C}_1$ carries also
a right-module structure, given by:
\begin{equation} \sum g_{ik}d_{ik}\cdot f:=\sum g_{ik}f_k\cdot d_{ik}\end{equation}
(For convenience we do not distinguish notationally between
elements of $\Fc$ and their Hilbert space
representations).

An important object in various areas of modern analysis on manifolds
or in Connes' approach to noncommutative geometry is the so-called {\it Dirac
  operator} $D$ (or rather, a certain version or variant of its
classical counterpart; for the wider context see
e.g. \cite{4} or \cite{Landi} to \cite{Esposito}). As $D$ we will take
in our context the operator:
\begin{equation} D:=\left( \begin{array}{cc}0 & d^{\ast}\\d & 0 \end{array} \right)\end{equation}
acting on
\begin{equation} H=\left( \begin{array}{c}H_0 \\ H_1 \end{array} \right)\end{equation}
with
\begin{equation} d^{\ast}=(\delta_1-\delta_2)\end{equation}
Note however, that there may exist in general several
possibilities to choose such an operator. On the other hand, we
consider our personal choice to be very natural from a geometrical
point of view.
\begin{lemma}There exists in our scheme a natural chirality- or
  grading operator, $\chi$ and an antilinear involution, $J$. given by
\begin{equation}\chi:=\left( \begin{array}{cc}1 & 0\\0 & -1 \end{array} \right)\end{equation}
with
\begin{equation}[\chi,{\Fc}]= 0    \quad \chi\cdot
  D+D\cdot\chi= 0\end{equation}
and
\begin{equation}J:\left( \begin{array}{c}x \\ y \end{array} \right)\to\left( \begin{array}{c}\overline{x} \\\overline{y} \end{array} \right)
\end{equation}
so that
\begin{equation}J\cdot f\cdot J=\overline{f}\end{equation}
\end{lemma}
These are some of the ingredients which establish what
Connes calls a \tit{spectral triple} (cf. e.g. \cite{Grav} or
\cite{Finite}). We do not want, however, to introduce the full
machinery at the moment as our scheme has an independent geometric
meaning of its own. So, being careful, we call in the following these
structures simply \tit{spectral triplets} (we were kindly warned by B.Iochum
to be more careful with this concept; note the observation below
about the non-compactness of the inverse of such Dirac-operators on
infinite graphs with a uniformly bounded vertex degree). 
\begin{defi}[Spectral Triplets] As {\em spectral
  triplet} on a general (undirected) graph we take 
\begin{equation} (H,{\Fc},D)\end{equation}
\end{defi}

At this point we would like to remark the following. In our general framework
we restricted ourselves, mostly for (possibly subjective) aesthetic
reasons -- the mathematics tends to be more transparent -- to
\tit{undirected graphs} and a total Hilbert space being the direct sum
of the \tit{node space} (a function space) and the \tit{bond space}
(sort of cotangent vectors). $A$ was then selfadjoint and a Dirac
operator emerged naturally as kind of a square root of the Laplacian.

On the other side, if one studies simple models as e.g. in \cite{5} to
\cite{7}, other choices are possible. In \cite{5},\cite{6}, where the
one-dimensional lattice was studied, the \tit{symmetric difference
  operator} was taken as Dirac operator. In \cite{7} the
one-dimensional lattice was assumed to be directed (i.e. only
$d_{i,i+1}$ were present) and the Dirac operator was defined as a
certain self adjoint ``doubling'' of the (one-sided,
i.e. non-symmetric) adjacency matrix. This latter model would fit in
our general approach if we had included more general graphs. All these
Dirac operators are different and it is hence no wonder that they lead
to different consequences (see below). It is our opinion that, in the end, an
appropriate choice has to be dictated by physical
intuition. Nevertheless, this apparent non-uniqueness should be
studied more carefully.

As can be seen from the above, the connection with the graph Laplacian
is relatively close since:
\begin{equation} D^2=\left( \begin{array}{cc}\dast d & 0\\0 & d\dast \end{array}
\right)\end{equation}
and
\begin{equation} \dast d=-2\Delta\end{equation}
$d\dast$ is the corresponding object on $H_1$. (In the vector analysis
of the continuum the
two entries correspond to $\operatorname{divgrad}\,
,\,\operatorname{graddiv}$ respectively ).
\begin{ob}Note that all our operators are bounded, the Hilbert space
  is (in general) infinite dimensional, hence there is no chance to
  have e.g. $(D-z)^{-1}$ or $(D^2-z)^{-1}$ compact. At the moment we
  are sceptical whether this latter phenomenon dissappears generically
  if the vertex degree is allowed to become infinite.  There are some
  results on spectra of random graphs which seem to have a certain
  bearing on this problem.
\end{ob}

We now calculate the commutator $[D,f]$ applied to an element $f'\in H_0$:
\begin{equation} (d\cdot f)f'=\sum_{ik}(f_kf'_k-f_if'_i)d_{ik}\end{equation}
\begin{equation} (f\cdot d)f'=\sum_{ik}f_i(f'_k-f'_i)d_{ik}\end{equation}
hence
\begin{equation} [D,f]f'=\sum_{ik}(f_k-f_i)f'_kd_{ik}\end{equation}
On the other side the right-module structure allows us to define $df$
as an operator on $H_0$ via:
\begin{equation} df\cdot f'=(\sum_{ik}(f_k-f_i)d_{ik})\cdot (\sum_k f'_kn_k)=\sum_{ik}(f_k-f_i)f'_kd_{ik}\end{equation}
In a next step we define $df$ as operator on $H_1$ which is not as
natural as on $H_0$. We define: 
\begin{equation} df|_{H_1}:\;d_{ik}\to (f_i-f_k)n_k\end{equation}
and linearly extended. A short calculation shows
\begin{equation}df|_{H_1}=-(d\bar{f}|_{H_0})^*=[d^*,f] \end{equation}
This then has the following desirable consequence:
\begin{ob} With the above definitions the representation of $df$ on
  $H$ is given by
\begin{equation}df|_H=\begin{pmatrix}0 & df|_{H_1} \\ df|_{H_0} &
    0\end{pmatrix}=\begin{pmatrix}0 & -(d\bar{f}|_{H_0})^* \\
    df|_{H_0} & 0\end{pmatrix}  \end{equation}
and it immediately follows
\begin{equation}df|_H=\begin{pmatrix}0 & [d^*,f] \\ {[}d,f] & 0\end{pmatrix}=  [D,f] \end{equation}
\end{ob}

\section{The Connes-Distance Function on Graphs}
From the general theory we know that:
\begin{equation}\label{eqqq} \|T\|=\|T^{\ast}\|\end{equation}
Hence
\begin{lemma}
\begin{equation}\label{eq} \|[d,f]\|= \|[d,\bar{f}]\|=\|[d^*,f]\|\end{equation}
and
\begin{equation} \|[D,f]\|=\|[d,f]\|\end{equation}
\end{lemma}
Proof: The left part of (\ref{eq}) is shown below and is a consequence
of formula (\ref{eqq}); the right identity follows from (\ref{eqqq}). With
\begin{equation} X:=\left( \begin{array}{c}x \\ y \end{array} \right)\end{equation} 
and $T_1:=[d,f]$,\;$T_2:=[d^*,f]$, the norm of $[D,f]$ is:
\begin{equation}
\|[D,f]\|^2=\sup\{\|T_1x\|^2+\|T_2y\|^2;\,\|x\|^2+\|y\|^2=1\}\end{equation}
Normalizing now $x,y$ to $\|x\|=\|y\|=1$ and representing a general
normalized vector $X$ as:
\begin{equation} X=\lambda x+\mu
y\;,\;\lambda,\mu>0\;\mbox{and}\;\lambda^2+\mu^2=1 \end{equation}
we get:
\begin{equation}
\|[D,f]\|^2=\sup\{\lambda^2\|T_1x\|^2+\mu^2\|T_2y\|^2;\|x\|=\|y\|=1,\lambda^2+\mu^2=1\}\end{equation}
where now $x,y$ can be varied independently of $\lambda,\mu$ in
their respective admissible sets, hence:

\begin{equation}
  \|[D,f]\|^2=\sup\{\lambda^2\|T_1\|^2+\mu^2\|T_2\|^2\}=\|T_1\|^2 \quad  
\Box \end{equation}
(as a consequence of equation (\ref{eq})).

It follows that in calculating $\|[D,f]\|$ one can restrict oneself to
the easier to handle $\|[d,f]\|$. For the latter expression we then get
from the above ($x\in H_0$):
\begin{equation}\label{eqq} \|df\cdot x\|^2=\sum_i(\sum_{k=1}^{v_i}|f_i-f_k|^2\cdot|x_i|^2\end{equation}
Abbreviating 
\begin{equation} \sum_{k=1}^{v_i}|f_k-f_i|^2=:a_i\geq 0\end{equation}
and calling the supremum over $i$ $a_s$, it follows:
\begin{equation} \|df\cdot x\|^2= a_s\cdot(\sum_i a_i/a_s\cdot|x_i|^2)\leq
a_s\end{equation}
for $\|x\|^2=\sum_i |x_i|^2=1$.

On the other side, choosing an appropriate sequence of normalized basis vectors $e_{\nu}$ so
that the corresponding $a_{\nu}$ converge to $a_s$ we get:
\begin{equation} \|df\cdot e_{\nu}\|^2\to a_s\end{equation}
We hence have
\begin{satz} \label{norm}
\begin{equation} \|[D,f]\|=\sup_i(\sum_{k=1}^{v_i}|f_k-f_i|^2)^{1/2}\end{equation}
\end{satz}

The \tit{Connes-distance functional} between two nodes, $n,n'$, is now defined as
follows:
\begin{defi}[Connes-distance function]
\begin{equation} dist_C (n,n'):=\sup\{|f_{n'}-f_n|;\|[D,f]\|=\|df\|\leq 1\}\end{equation}
\end{defi}
\begin{bem} It is easy to prove that this defines a metric on the
graph.
\end{bem}
\begin{koro} It is sufficient to vary only over the set
$\{f;\|df\|=1\}$.
\end{koro}
Proof: This follows from
\begin{equation} |f_k-f_i|=c\cdot|f_k/c-f_i/c|\;;\;c=\|df\|\end{equation}
and
\begin{equation} \|d(f/c)\|=c^{-1}\|df\|=1\end{equation}
with $c\leq1$ in our case.\bewende
\vspace{0.5cm}

It turns out to be a nontrivial task (in general) to calculate this
distance on an arbitrary graph as the above constraint is quite subtle
. The underlying reason is that the constraint is, in some sense,
inherently \tit{non-local}. As $f$ is a function, $f_{n'}-f_n$ has to be
the same independently of the path connecting $n'$ and $n$. On the
other side, in a typical optimization process one deals with the
individual jumps, $f_k-f_i$, along some path. It is then not at all
clear that these special choices can be extended to a global function
without violating the overall constraint on the expression in theorem
\ref{norm}. Nevertheless we think the above closed form is a solid
starting point for the calculation of $dist_C$ on various classes of
graphs or lattices. We discuss two examples below but refrain at this
place from a more complete treatment, adding only some observations
concerning the relation to the ordinary (combinatorial) distance
function introduced in the beginning of the paper.

Having an admissible function $f$ so that
$\sup_i(\sum_{k=1}^{v_i}|f_k-f_i|^2)^{1/2}\leq 1$, this implies that,
taking a {\it minimal path} $\gamma$ from, say, $n$ to $n'$, the
jumps $|f_{\nu+1}-f_{\nu}|$ between neighboring nodes along the path
have to fulfill:
\begin{equation} |f_{\nu+1}-f_{\nu}|\leq 1\end{equation}
and are typically strictly smaller than $1$ as long as there are not a
sufficient number of ``zero-jumps'' ending at the same node.

On the other side the Connes distance would only become identical to the
ordinary distance $d(n,n')$ if there exist a sequence of admissible
node functions with all these jumps approaching the value $1$ along such a
path, which is however impossible in general as can be seen from the
structure of the constraint on the expression in theorem \ref{norm} . Only in this case
one may have a chance to get:
\begin{equation}| \sum_{\gamma}(f_{\nu+1}-f_{\nu})|\to
\sum_{\gamma}1=length(\gamma)\end{equation}
We express this observation in the following way
\begin{ob}[Connes-distance] One has within our general scheme the following
inequality
\begin{equation} dist_C(n,n')\leq d(n,n')\end{equation}
A fortiori one can prove that $dist_C$ between two nodes in
an arbitrary graph is even smaller than or equal to the corresponding
Connes-distance taken with respect to the (one-dimensional) sub-graph
formed by a minimal path between these nodes, i.e.
\begin{equation}dist_C(n,n')\leq dist_C(min.path)(n,n')\end{equation}
The simple reason is that one has more \tit{admissible functions} at
ones disposal for a subgraph, hence the supremum may become larger.
This latter distance, on the other side, can be rigorously calculated (see Example 2
below) and is for non-neighboring nodes markedly smaller than the
ordinary distance.
\end{ob}
\begin{koro}The last inequality implies also that with $G'$ an induced
  subgraph of $G$ it holds ($n,n'\in V'\subset V$):
\begin{equation}dist_C(n,n';G')\leq dist_C(n,n';G)\end{equation}
\end{koro}

We remarked above that the calculation of the Connes distance on
graphs is to a large part a continuation problem for admissible
functions, defined on subgraphs. Then the following question poses itself.
\begin{problem}For what classes of graphs and/or subgraphs do we have
  an equality in the above corollary?
\end{problem}
\begin{bem}Equality can e.g. be achieved for trees. Other results in
  this direction are in preparation.
\end{bem}

These general results should be contrasted with the results in \cite{5}
to \cite{7}. Choosing e.g. the symmetric difference operator as Dirac
operator in the case of the one-dimensional lattice the authors got in
\cite{5,6} a distance which is \tit{strictly greater} than the
ordinary distance but their choice does not fulfill the above natural
constraint given in Theorem \ref{norm}. Note in particular that our
operator $d$ is a map from node- to bond-functions which is not the
case in the other examples. In \cite{7} the authors employed a
symmetric doubling of the upper half of our symmetric adjacency matrix
as Dirac operator. In the case of the one-dimensional (directed)
lattice this then leads (so to say) to only one (directed) bond per node
and makes the optimization process quite simple, hence leading to the
ordinary distance which would have been also the case in our general
scheme had we admitted directed graphs. We conjecture however that for
more general graphs a relation related to the one given in
Theorem\ref{norm} would enforce the Connes-distance to be again
strictly smaller than the ordinary distance for non-neighboring
points. This is however an interesting point and we plan to discuss
generalisations of our framework and more general examples elsewhere.

We want to close this paper with the discussion of two examples. The
first one is a simple warm-up exercise, the second one is the
one-dimensional lattice discussed also by the other authors mentioned
above (treated however within their respective schemes) and is not so
simple. The technique used in approaching the second problem may be
interseting in general. While we solved it starting, so to speak, from
first principles, the real mathematical context, to which the strategy
is belonging, is the field of \tit{(non-)linear programming} or
\tit{optimization} (see e.g. \cite{Jungnickel} or any other related
textbook). This can be inferred from the structure of our constraint
on the expression in theorem \ref{norm}. This means that the techniques developed in
this field may be of use in solving such quite intricate problems. 
\\[0.5cm]
\tit{Example 1: The square with vertices and edges}:
\begin{equation} x_1-x_2-x_3-x_4-x_1\end{equation}
Let us calculate the Connes-distance between $x_1$ and $x_3$. 
As the $\sup$ is taken over functions(!) the summation over elementary
jumps is (or rather: has to be) pathindependent (this is in fact both a subtle and crucial
constraint for practical calculations). It is an easy exercise to see
that the $sup$ can be found in the class where the two paths between
$x_1,x_3$ have the {\it valuations} ($1\geq a\geq0$):
\begin{equation} x_1-x_2:\,a\;,\;x_2-x_3:\,(1-a^2)^{1/2}\end{equation}
\begin{equation} x_1-x_4:\,(1-a^2)^{1/2}\;,\;x_4-x_3:\,a\end{equation}
Hence one has to find $\sup_{0\leq a\leq1}(a+\sqrt{1-a^2})$. Setting
the derivative with respect to $a$ to zero one gets
$a=\sqrt{1/2}$. Hence:\\[0.5cm]
\begin{bsp}[Connes-distance on a square]
\begin{equation} dist_C(x_1,x_3)=\sqrt{2}<2=d(x_1,x_3)\end{equation}
\end{bsp}
\tit{Example 2: The undirected one-dimensional lattice}:\\[0.5cm]
The nodes are numbered by $\Z$. We want to calculate $dist_C(0,n)$
within our general framework. The calculation will be done in two main
steps. In the first part we make the (in principle quite complicated)
optimization process more accessible. For the sake of brevity we state
without proof that it is sufficient to discuss real monotonely
increasing functions with
\begin{equation}f(k)=\begin{cases}f(0) & \text{for}\;k\leq 0\\f(n) &
    \text{for}\;k\geq n \end{cases}
\end{equation}
and we write
\begin{equation}f(k)=f(0)+\sum_{i=1}^k h_i\quad\text{for}\quad 0\leq
  k\leq n\;h_i\geq 0\end{equation}
The above optimization process then reads:
\begin{ob}Find $\sup \sum_{i=1}^n h_i$ under the constraint
\begin{equation}h_1^2\leq 1,h_2^2+h_1^2\leq
  1,\ldots,h_n^2+h_{n-1}^2\leq 1,h_n^2\leq 1\end{equation}
\end{ob}
The simplifying idea is now the following. Let $h:=(h_i)_{i=1}^n$ be
an admissible sequence with \tit{all} $h_{i+1}^2+h_i^2<1$. We can then
find another admissible sequence $h'$ with
\begin{equation}\sum h_i'>\sum h_i\end{equation}
Hence the supremum cannot be taken on the interior. We conclude that
at least some $h_{i+1}^2+h_i^2$ have to be one. There is then a
minimal $i$ for which this holds. We can convince ourselves that the
process can now be repeated for the substring ending at $i+1$.
Repeating the argument we can fill up all the entries up to place
$i+1$ with the condition $h_{l+1}^2+h_l^2=1$ and proceeding now
upwards we end up with
\begin{lemma}The above supremum is assumed within the subset
\begin{equation}h_1^2\leq
  1,h_1^2+h_2^2=1,\ldots,h_{n-1}^2+h_n^2=1,h_n^2\leq 1\end{equation}
\end{lemma}
This concludes the first step.

In the second step we calculate $\sup|f(0)-f(n)|$ on this restricted
set. From the above we now have the constraint:
\begin{equation}h_1^2\leq
  1,h_2^2=1-h_1^2,h_3^2=h_1^2,h_4^2=1-h_1^2,\ldots,h_n^2=1-h_1^2\;\text{or}\;h_1^2 \end{equation}
depending on $n$ being even or uneven. This yields
\begin{equation}\sup|f(0)-f(n)|=\begin{cases}
1 & \text{for}\;n=1\\
(n/2)\cdot\sup(h_1+\sqrt{1-h_1^2})=(n/2)\cdot\sqrt{2} &
\text{for $n$ even}\\ 
\sup ({[}n/2]\cdot(h_1+\sqrt{1-h_1^2})+h_1) & \text{for $n$ uneven}
\end{cases}
\end{equation}
In the even case the rhs can be written as
$\sqrt{n^2/2}=\sqrt{[n^2/2]}$. In the uneven case we get by
differentiating the rhs and setting it to zero:
\begin{equation}h_1^{max}=A_n/\sqrt{1+A_n^2}\;,\;\sqrt{1-(h_1^{max})^2}=1/\sqrt{1+A_n^2}\end{equation}
with $A_n=1+1/[n/2]$. We see that for increasing $n$ both terms
approach $1/\sqrt{2}$, the result in the even case. Furthermore we see that the
distance is monotonely increasing with $n$ as should be the case for a
distance. This yields in the uneven case
\begin{equation}\label{uneven} dist_C(0,n)=\frac{([n/2]+1)A_n+[n/2]}{\sqrt{1+A_n^2}}\end{equation}
which is a little bit nasty. Both expressions can however be written in a
more elegant and unified way (this was a conjecture by W.Kunhardt,
inferred from numerical examples). For $n$ uneven a short calculation yields
\begin{equation}[n^2/2]=(n^2-1)/2=1/2\cdot(n-1)(n+1)=2[n/2]([n/2]+1)\end{equation}
(with the \tit{floor-,ceiling-}notation the expressions would become
even more elegant). With the help of the latter formula the rhs in
(\ref{uneven}) can be transformed into
\begin{equation}rhs\;\text{of}\;(\ref{uneven})=\sqrt{[n^2/2]+1}\end{equation}
\begin{conclusion}For the one-dimensional undirected lattice we have
\begin{equation}dist_C(0,n)=\begin{cases} \sqrt{[n^2/2]} & \text{for
      $n$ even}\\\sqrt{[n^2/2]+1} & \text{for $n$ uneven}
\end{cases}
\end{equation}
\end{conclusion}
Remark: With the help of the methods, introduced above, we can now
estimate or rigorously calculate the Connes-distance for other classes
of graphs.\\[0.5cm]
Acknowledgement: We thank the referees for their constructive criticism.

\end{document}